# Extremely Large Magnetic Entropy Changes, Quantum Phases, Transitions and Diagram in Gd(OH)$_3$ Single Crystal Nanowires - Quasi-1D Large Spin ($S = -7/2$) Chain Antiferromagnet

R. Zeng [1, 2], C.F. Zhang[1], J.C. Debnath[1], P. Shamba[1], J.L. Wang[1], Z.X. Chen[1], Z.P. Guo[1], S.X. Dou[1]

[1]*Institute for Superconducting and Electronic Materials, University of Wollongong, NSW 2522, Australia*

[2]*School of Computing, Engineering and Mathematics, University of Western Sydney, Campbelltown, NSW 2560, Australia*

Address for Correspondence:

R. Zeng

Solar Energy Technologies
School of Computing, Engineering and Mathematics
University of Western Sydney
Penrith Sout, Sydney, NSW 2751, Australia
Electronic mail: r.zeng@uws.edu.au, or rzeng@uow.edu.au.

# Abstract

The study of quantum phase transitions (*QPT*) has continually attracted great interest from many researchers over the last three decades [1-7], especially since the discovery of the cuprate high temperature superconductors. The magnetocaloric effect in quantum spin systems has recently attracted renewed attention because field-induced *QPT* leads to universal responses when the applied field is varied adiabatically [3, 8], and the magnetocaloric effect (*MCE*) is the only way to approach the limit of low temperature ($T \rightarrow 0$ K). In terms of applications of MCE, magnetic refrigeration (MR) has become a promising alternative to conventional refrigeration technology [9-14] due to its more energy-efficient and environmentally-friendly features, and the exotic paramagnetic (PM) phases and transitions that accompany it have become more and more important for fundamental research, quantum computing, and space technology application [15-18]. The MCE is defined as the temperature change in magnetic materials as a result of alignment of spins when the material is exposed to a magnetic field. Giant MCE near room temperature has been reported in recent years, but there are still problems with hysteresis (due to the nature of first order ferromagnetic phase transitions) for practical application[13]. At the same time, high MCE materials with molecular clusters and nanostructures for access to ultra-low temperature have also been proposed and successfully measured and confirmed [15-28], but the MCE values still seem somewhat low for high MR efficiency. On the other hand, low temperature MCE or even zero-point entropy study is important for our understanding of complex phase structures and exotic paramagnetic phase transitions of many low dimensional and frustrated magnetic materials. Here, we have conducted systematically magnetic and magnetothermal measurements at temperatures down to 2 K and magnetic fields up to 13.5 Tesla, analysis and report that, (1) magnetic field enhances the thermal and local spin fluctuations which suppress long-range spin ordering (LRO) within the measured temperature range, and close to 0 K at the quantum critical point (QCP); (2) possible field-induced exotic local spin-liquid-like, aligned-spin, and spin-flip exotic paramagnetic phases, and transitions in the low temperature and high field range have been observed, allowing us to identify a possible quantum critical point; (3) there is extremely large, fully reversible MCE (magnetic entropy change ($-\Delta S_M$) = 27.8, 66, and 88 J / kg K, adiabatic temperature change ($\Delta T_{ad}$) = 6.7, 17.6, and 36.4 K at 2.55 K for field changes of 2, 5, and 11 T, respectively in the continuum of quantum phase transitions in this system; (4) moreover, careful experiments and analysis may allow experimental determination and set up a quantum phase diagram of this system. The magnetic-entropy change monotonically increases with decreasing temperature, and it exceeds the magnetocaloric effect (MCE) in any other known low temperature reversible MCE material by at least a factor of 3. The extremely large magnetic entropy change may be attributed to the large amount of weakly interacting spins that can be easily aligned at low-lying energy in the quantum critical regime of our nanosized materials, since there is large MCE in the local spin-liquid-like (low energy excitation and even gapless state) range. These indicate that the material is a promising MCE candidate for low temperature application, and possibly could make ultra-low temperatures easily achievable for most laboratories and for space application as well.

## Introduction

A large magnetocaloric effect (MCE) can be obtained near the magnetic-ordering temperature by aligning the spins under application of an external magnetic field. A giant MCE is usually found to be related to a field-induced first-order magnetic transition [9-12]. However, a first-order transition usually gives rise to considerable thermal/magnetic hysteresis due to concurrent field-induced crystallographic structural changes, which is disadvantageous for applications. Therefore, much attention has been focused on finding new materials with a large MCE and small thermal/magnetic hysteresis [5, 6]. Research is in progress to find new materials which have large MCE at low fields close to room temperature for domestic and other technological applications. On the other hand, systems showing large MCE in the low-temperature region from about 20 K down to the sub-Kelvin range are also important for basic research, as well as for specific technological applications, such as space science and liquefaction of hydrogen and natural gas in the fuel industry [15 - 28]. In particular, when a temperature approaching absolute zero is needed ($T \rightarrow 0$ K), it is only magnetic refrigeration technology that can achieve it. On other hand, when the temperature $T \rightarrow 0$ K, according to thermodynamic principles, all matters should be ordered and show very low entropy and entropy change, while for some materials, especially for low dimensional magnetostructural magnets, the enhanced spin fluctuations would suppress magnetic ordering and give rise to a variety of exotic phenomena and novel types of magnetism, such as quantum critical phenomena and quantum phase transitions under changes in external conditions. One such example is the magnetic field induced quantum critical point. When the system emerges from the "local spin liquid" state, the system is in a highly degenerate and low-lying energy state. All the spins are singlets, but in strongly coherent, low-lying, even gapless energy states, so that external field can easy align them and translate the state to an aligned spin state, which may generate extremely high magnetic entropy change.

The quantum phase transitions (QPTs), in contrast to their classical counterparts at $T > 0$ K, where thermal fluctuations are important, are driven by a control parameter other than temperature, e.g. composition, pressure, or field. A quantum critical point (QCP) commonly separates an ordered from a disordered phase at $T \rightarrow 0$ K. The magnetocaloric effect may be enhanced by magnetic ion clusters [19] and magneto-structural geometric frustration [28], promising an improved efficiency in $T \rightarrow 0$ K refrigeration application. The development of magnetic refrigerant materials for ultra-low temperatures has mainly been focused on the garnets and paramagnetic salts [20]. However, those materials normally also present a somewhat high ordering temperature, which introduces a very low MCE at ultra-low temperature and hence limits the lowest temperatures achievable with paramagnetic salts. In recent years, superparamagnetic nanostructured materials [21, 22], well designed molecular nanomagnets and magnetic clusters [23-27], and frustrated magnets [19] have been introduced to improve the MCE required to reach $T \rightarrow 0$ K.

Here we introduce a new nanosized Gd(OH)$_3$ material, which shows a large spin ($S = 7/2$) quasi-one-dimensional (1D) chain with weak antiferromagnetic interaction [29]. We find that strong thermal and spin fluctuation takes place at low temperature, and long-range spin ordering does not exist within the measured temperature range, but we found that field-induced, spin-liquid-like (local spin-liquid), aligned-spin and spin-flip exotic paramagnetic phases and their transitions occur in the low temperature and high field range, making it possible to determine the quantum critical point and construct the phase diagram. Moreover, we have demonstrated an extremely large reversible magnetic-entropy change observed in the paramagnetic phase transition range of high spin Ising chains in Gd(OH)$_3$ single crystal nanowires. These indicate that, in contrast to classical phase transition, giant MCE can be obtained near the first order transition temperature, extremely large MCE can be obtained around the spin fluctuation temperature range, and in the continuous quantum phase transition temperature range.

## Abstract

Gd(OH)$_3$ nanowires were synthesized by a facile hydrothermal method, as described in the experimental methods section. Microstructure analysis indicate that the sample present highly purity and uniformity features, the details see supporting information (SI) in Fig. S1. The crystal structure dominates the magnetic properties of bulk Gd(OH)$_3$, with a weak nearest-neighbor isotropic exchange aligned in the *c*-direction, causing an antiferromagnetic interaction, and a much weaker next-nearest neighbor exchange that gives rise to a ferromagnetic interaction, plus a likely dipole interaction, which drives the long-range order to very low temperature (Néel temperature, $T_N \approx 0.94$ K) [29, 30].

The form of the effective spin Hamiltonian for Gd(OH)$_3$ should be identical to that of other isostructural Gd$^{3+}$ compounds, several of which have previously been studied in some detail [31, 32]. This has been calculated [29] according to experimental magnetic and magnetothermal measurement data and is expressed by the following spin Hamiltonian equation (1):

$$H_{ami} = \sum_{i>j} J_{ij} S_i S_j + \sum_{i>j} \alpha_{ij} [S_i S_j - 3 (S_i r_{ij})(S_j r_{ij})] + \sum_i g\mu_B H S_i + V_C - H \sum_i S_i \qquad (1)$$

where $\alpha_{ij} = g^2 \mu_B^2 / r_{ij}^2$ is a measure of the strength of the magnetic dipole-dipole coupling and $r_{ij}$ is the distance between ions *i* and *j*; the effective spin $S = 7/2$, the *g* factor has been found to be about 2, the term $V_C$ denotes the effect of the crystal field and indicates that the largest single interaction is the nearest-neighbor isotropic exchange $J_1 = 0.180$ K, which tends to weakly align nearest neighbors antiferromagnetically. The next-nearest-neighbor exchange is ferromagnetic, but much smaller, $J_2 = 0.017$ K, and the third nearest-neighbor exchange, in addition to the effect of the crystal field and all the sources of anisotropic-exchange and dipole interaction, is negligible in comparison. Problems arise when the largest interaction is not sufficient by itself to produce long-range order. In such cases, competition between several of the weaker interactions may then result in a relatively complex situation. Thus, the structure can be described as quasi-one-dimensional (*1D*) spin chain, and in such a system, there is a tendency towards instability, as the nanoscale structure makes the system more unstable, so that under external magnetic field *H*, there may be suppression of the long-range ordering temperature towards absolute zero (T → 0 K). The Hamiltonian equation can be simply described by the following Eq. (2) for this quasi-1D spin chain:

$$H_{ami} = \sum_{i>j} J_{ij}S_iS_j - H \sum_i S_i \quad (2)$$

The 1D Ising chain in transverse field is perhaps the most-studied theoretical paradigm for a quantum phase transition [1-6]. We cannot reach sub-Kelvin temperatures in our laboratory, however, our high field magnetic measurements and the magnetocaloric properties (down to 2 K) can be employed to predict the presence of any state with field-induced orderings, as Ref. [33] has demonstrated that some remarkable properties of the 1D quantum chain may have important implications for experimental characterization of QPTs when one is unable to reach temperatures below which a QPT can be seen. As well, proximity to a quantum critical point may be the cause of unusual properties in a variety of materials, even when the quantum critical point itself has not been observed. In such materials, the quantum critical point might only be observable with an unphysical tuning parameter such as negative pressure. Yet the proximity to quantum criticality would still leave a shadow on the phase diagram of the material, since the QCP is entirely an extrapolation of experimental data. Although the $Gd(OH)_3$ single crystal nanowire material is not a large single crystal, in the randomly aligned powder nanowires, part of the single crystal wires are under external transverse magnetic field when there is an applied field; in addition, although the magnetic anisotropy of $Gd(OH)_3$ material is not very large, reasonable anisotropy has been demonstrated in a single crystal sample [29]; in particular, the very low exchange energy $J$ in $Gd(OH)_3$ materials can be easy matched by experimentally attainable magnetic fields (10 T ≈ 1 meV) to access the QCP, so that high field measurements may be very beneficial to explore the exotic paramagnetic phases (or quantum phase) and QPTs, and to establish the quantum phase diagram.

A series of magnetic and magnetothermal measurements have been conducted to identify the possible magnetic states and possible quantum phase transitions. Detail magnetic susceptibility ($\chi$) analysis was shown in SI information (Fig. S2). The temperature dependence of the magnetic susceptibility ($\chi$) under zero field cooling (ZFC) and field cooling (FC) in an external magnetic field $H$ = 1000 Oe is plotted in Fig. S2(a). The curves can be well fitted as Curie-Weiss law behaviour over the whole temperature range, with the Curie-Weiss temperature, $\Theta_{CW} \approx -5.91$ K, the Curie constant, $C$ = 7.16 emu / mol, and the effective magnetic moment, $\mu_{eff}$ = 7.56 $\mu_B$, where $\mu_B$ is the Bohr magneton. The fitting line is shown by the solid blue line in Fig. S2(a). However, there is no spontaneous magnetic ordering transition that takes place within the temperature range from 2 to 300 K in this nanowire system. The negative value of $\Theta_{CW}$ is expected because of the potential for antiferromagnetic interaction below 2 K. We note that $\mu_{eff}$ = 7.56 $\mu_B$ is slightly lower than, but very close to the theoretical effective magnetic moment of non-interacting $Gd^{3+}$ ions ($\mu_{eff}$ = 7.95 $\mu_B$) [32]. We also note that the ZFC and FC susceptibility curves are the same, and there is no so-called superparamagnetic interaction behaviour that would cause a blocking temperature to appear in the 2 – 300 K temperature range, which indicates that all the interaction or spin ordering potential for long-range ordering is suppressed to ultra-low temperature (< 2 K). In thermodynamic measurements, $\Theta_{CW}$ is the expected ordering temperature in the absence of frustration [34], suggesting a simple figure of the frustration constant – the frustration factor ($f$) [34], $f = \Theta_{CW} / T_C(T_N)$, for measuring the influence of frustration on the cooperative transition. In this Gd(OH)$_3$ nanowire system, $f > 6$, since the $T_N$ should be lower than 0.94 K [29], as the $T_N$ in a nanoscale material is lower than in the corresponding bulk system. That $1/\chi$ is linear at such a small value of $T_N/\Theta_{CW}$ is surprising in itself and represents another characteristic feature of frustrated systems. In non-frustrated systems, $\chi$ starts to deviate from Curie-Weiss behaviour at $T \approx 2\Theta_{CW}$, making a linear extrapolation of $1/\chi$ possible only if there is data over a significant range of temperature above $2\Theta_{CW}$. Since this deviation from Curie-Weiss behaviour is a result of growing correlations among the spins, e.g., most likely due to the Ruderman-Kittel-Kasuya-Yosida (*RKKY*) interactions of the *4f*$^7$ unpaired electrons moment for the negligible of (i) superexchange and dipolar interaction in the insulating 1D Ising chain spin structure, (ii) single-ion Kondo coupling, and (iii) crystal electric field (*CEF*) scheme of $Gd^{3+}$. Since mean-field theory is inadequate for predicting the critical temperature in large frustration factor *f* value systems, it is inapplicable for describing the linearity of $1/\chi$. The value of $f > 6$ means that this Gd(OH)$_3$ spin chain is an intermediate frustrated system, and the unusual features of the above-mentioned remarkably simple behaviour can be currently explained by a local spin liquid-like scenario. [35-37] Therefore, the strongly frustrated systems have the unusual feature that the spins behave like free entities, despite the extremely large temperature-renormalized coupling strength. This means that the system may emerge from a local spin liquid-like state under finite conditions.

Those results suggest that the system is a simple spontaneous paramagnet and that superparamagnetic behaviour does not appear in the temperature range from 2 to 300 K, but typically, exotic paramagnetic phases and complex transitions might be observed due to the weakly frustrated quasi-1D spin chain structure, where strong thermal and local spin fluctuations compete with the spontaneous antiferromagnetic interaction. We indeed observed that the relationship between susceptibility ($\chi$) and temperature shows significant changes under high field, as shown in Fig. S2(b), where the where $\chi$ - $T$ curves under field from 1 kOe to 135 kOe are presented, and the inset shows the $\chi$ vs. log $T$ curves. We note that there is a divergence in the susceptibility ($\chi$) for $T \to 0$ K that appears in the magnetic field range from 1 to 2 T, and a critical field $H^*_C$ may be present around 1-2 T. As shown in Fig. S2(b) and the $1/\chi$–$T$ curves shown in Fig. 1(a), when $H < H^*_C$, the curves can be well fitted to the Curie-Weiss (CW) rule – the system is strictly a CW paramagnet, but when $H > H^*_C$, the curves show a downturn ($\chi$ - $T$ curves) or upturn ($1/\chi$–$T$ curves). A systematic study has revealed a divergence in the susceptibility ($\chi$) for $T \to 0$ K at a critical point. The spontaneous moment vanishes linearly at the critical point [37]. We employed a similar method to analyse the critical phenomena and critical points at $T \to 0$ K in our system. At $H < H^*_C$, the system falls into the classical conventional range, and we use the Curie-Weiss (CW) rule to analyse the system and determine the spontaneous inverse susceptibility $1/\chi_o$ at $T \to 0$ K. At $H > H^*_C$, the system falls into the quantum phase transition regime, and we directly extrapolate from the experimental data to determine the inverse susceptibility $1/\chi_o$ at $T \to 0$ K. Fig. 1(b) clearly shows that the $|1/\chi_o|$ value linearly decreases with the logarithm of the external magnetic field value in the $H < H^*_C$ range. This indicates, as is understandable, that the field suppresses antiferromagnetic interaction due to the field enhanced spin fluctuation. When a line is fitted to the curve, there is a relationship of $|1/\chi_o| = 86.225 - 12.5 \log (\mu_0 H)$. At the quantum critical point, $T \rightarrow 0$ K, where $\chi_o \rightarrow \infty$, $|1/\chi_o| \rightarrow 0$, we obtain the quantum critical field, $H^*_C \approx 1.4$ T. The two inset figures in Fig. 1(b) show the determination of the $1/\chi_o$ value at $H < H^*_C$ (left inset) and at $H > H^*_C$ (right inset), respectively.

Moreover, we noted that the shape of the curves is similar to what appears for the spin disordered triangular lattice in NiGa$_2$S$_4$ [38], where the liquid-like state is stabilised by geometrical frustration in this system. In our system, clear crossover behaviour is seen for the log $(\chi)$ versus log $(T)$ curves, which are characterized by inflection points [39] (blue arrows shown in inset of Fig. S2(b)), where the points are denoted as T*. It is clear from the figure that T* increases with increasing field H, and when $T > T^*$, the $\chi$ - T curves are well fitted by the Curie-Weiss (CW) rule – the system is strictly a CW paramagnet: $\chi \propto 1/T$; when $T < T^*$, the relationship is $\chi \propto - \log (T)$. The boundary position line at different external fields can be obtained from the crossover transition points in Fig. S2(b). Another point, denoted as $T_C^*$, shown in Fig. S2(b), is the crossover point of log $(\chi) \approx$ constant A with the CW straight line, and it shows that the value of A only depends on the external field. The $1/\chi$–T curves, as shown in Fig. 2(a), show the characteristic more clearly. In the low temperature range, at the upturn point, the temperature increases with increasing field. The inset of Fig. 1(a) shows the $1/\chi$–T curve under 135 kOe field, indicating that at extremely low temperature, the curve shows another more significant upturn: this point is the spin-flip temperature that we have denoted as $T_{SF}$, which will be discussed later. In order to clearly pick out the critical point temperature, we replotted the curves as $\delta(1/\chi)/\delta T - T$ curves, as shown in Fig. 1(b). We denoted the peak point as $T_{AS}$, as it is the temperature where the spin state becomes aligned under external field. For example, $T_{AS}$ =15.4 K and $T_{SF}$ = 3.6 K under field of 135 kOe, as shown in Fig. S2(c), and these values are confirmed by the magnetisation curves below and the thermodynamic properties measurements. For the spin-liquid-like model, we observed a sharp decrease in the $\chi T$ value, as shown in Fig. S2(e). The $\chi(T) T$ value $\equiv$ effective Curie constant $C(T)$, e.g., $\chi(T) T \equiv C(T)$, while there is a linear decrease in $C(T)$ at $H \geq 2T$ with the same slope of 0.0144, and the relationship appears to be $\chi(T) T \equiv C(T) \propto 0.0144 \log(T)$ or $\chi(T) \propto 0.0144 \, T^{-1} \log(T)$. The magnetic field only varies the temperature range where the $C(T)$ ceases to be linear and is reduced with decreasing temperature. The abnormal linear relationship of $C(T)$ or $\chi(T)$ to temperature suggests spin-liquid-like short-range correlations, which may due to (i) the RKKY long-range interaction, which is compatible with an antiferromagnetic interaction; (ii) strong spin and thermal fluctuations. Although there is weak next-nearest-neighbor exchange, the frustration factor $f > 6$ indicates that the system falls into the intermediate frustration spin structure range, which may due to the slight structural disorder on the nanoscale f, and also, the $\chi T$ value around 26 K to 10 K appears to fluctuate in a field of 1 kOe, as shown in the inset to Fig. S2(e), which suggests that strong spin and thermal fluctuations may take place in this temperature range and that the fluctuation starting temperature increases with increasing applied field, as shown in Fig. S2(c) and S2(e). Fig. S2(g) shows the temperature dependence of $\chi T$ under an external magnetic field of 50 kOe, and it clearly illustrates the exotic phases and their changes from strong fluctuations to a spin-liquid-like state, and then to aligned spin states. In fact, the spin structure and magnetism featured in this system is similar to the dimensional reduction system of $Cs_2CuCl_4$, with a spin -1/2 antiferromagnet on a moderately anisotropic triangular lattice and intermediate frustration ($f = 8$) ordering into a spiral at $T_N = 0.6$ K [40-42]. Although the interchain exchange coupling is weaker, an order of magnitude smaller than for this similar $Ga(OH)_3$ system, the quantum magnetism is extremely sensitive to the interchain exchange coupling at low temperature [43]. So, we can consider this system as consisting of quasi-one-dimensional intermediate frustrated spin 7/2 antiferromagnetic chains. Moreover, the system emerges from the spin-liquid-like state only at infinite range (as shown in Fig. S2(e) and S(g)), which indicates that other interactions can not be neglected and that where Eq. (2) is too simple to interpret the system, Eq. (1) should be used.

In order to cross-check the exotic paramagnetic phases and their transitions, a series of isothermal $M – H$ curves (as shown in Fig. S3(a)) has been collected, and they show a change in curvature from the more linear shape at high temperatures to more bent shapes at lower temperatures (meaning ferromagnetic-like (FM-like) behaviour, but still in the paramagnetic (PM) state) when the temperature is lower than 25 K, which indirectly supports the existence of the local spin fluctuations. Fig. S3(a) shows the FM-like state, but the Arrott plots of the $M – H$ curves (as shown in Fig. S3(b)) don't show the classical magnetic phase transition characteristic, e.g., the Arrott plot should show a linear relationship at the transition temperature, but in fact, it shows a convex curvature against $H/M$ in the low temperature range. To the best of our knowledge, the convex curvature of $M^2$ against $H/M$ has so far been reproduced only by Takahashi's spin-fluctuation theory [44, 45]. Moreover, we plotted the isothermal magnetization curves in the modified Arrott plot form of $M^4$ vs. $H/M$, as shown in Fig. S3(c). Here, the $M^4$ shows an almost linear relationship to $H/M$ around the temperature range of 10 to 25 K, which confirms the above mentioned spin fluctuation range shown in Fig. S2(c-g). According to Takahashi's theory, in the case where the coefficient of $M^4$ in the Landau expansion of the free energy is zero, $M^4$ becomes linear against $H/M$. Thus, complete linearity is only obtained under delicate conditions. The system appears to be very near those conditions. We note that the maximum value of the moment at the temperature of 2.1 K reaches $M_{max} = 7.14$ $\mu_B/Gd^{3+}$, and this value is close to the theoretical value of $Gd^{3+}$, where almost every single spin is aligned at high field and the system seems to have fallen into the aligned spin state (FM-like). The aligned spin in a high energy state is not stable, and with increasing field, the system would be transformed into other states with lower energy. One of the most likely types of behavior is the formation of a resonating-valence-bond (RVB) state [46, 47] by spin flipping. We also note that the spin flipping leads to the emergence of $E_8$ symmetry in the $CoNb_2O_6$ Ising chain [48]. We have observed this striking phenomenon in the high field range in this quasi-1D spin chain system.

In order achieve a more detailed understanding of the low temperature and high field spin state, we differentiated the $M – H$ curves and have plotted them in Fig. S3(c). At low field, where $\delta(M)/\delta H \approx$ constant, the system is in the paramagnetic state; in the middle field range, the system makes a transition from the paramagnetic state to short range spin order or a spin liquid, and then to a highly ordered aligned spin state in a continuous process with increasing field. It is also interesting to note that there is a crossover point at a field of about 1.4 T for the $\delta M/\delta H – H$ curves at the indicated temperature, which we have denoted as $H_C^* = 1.4$ $T$. More detailed $M – H$ curves at temperatures from 2.1 K to 4.1 K are shown in Fig. 2(a), where the left inset shows a magnified view of the low field region, and the right inset shows a magnified view of the high field region. The low field $M – H$ curves are linearly increasing with the field, but there are complex relationships (an emerging spin liquid and aligned spin state) in the middle field area. It is interesting to note that the $M – H$ curves reach a peak of maximum moment and then linearly decrease with increasing field. This is actually due to the field induced spin flip state. We have denoted the field as $H_{SF}$ at indicated temperatures $T_{SF}$, as shown in the lower right inset of Fig. 2(a), and we thus obtain a $H_{SF}$-$T_{SF}$ boundary line for the phase diagram. Fitting the straight line parts of the $M – H$ curves at low field (upper left inset of Fig. 2(a)) and high field (lower right inset of Fig. 2(a)), the slopes are the spin susceptibilities ($\chi_M = \delta M/\delta H$) at different temperatures. As for the emerging spin flip field ($H_{SF}$) at the indicated temperatures, the results at indicated temperatures $T$ and field $H_{SF}$ are plotted in Fig. 2(b). Fitting the low field $\chi_M – T$ lines and the high field $\chi_M – T$ and $\chi_M – H_{SF}$ lines, as shown in Fig. 2(b), equations are obtained for both the low and high field ranges with

$$\chi_M = 13.1 – T^{0.7} \ (10^{-3} \text{ emu/g Oe}) \quad \text{(low field range)} \quad (3)$$
$$\chi_M = – 2.8 + T^{4.5} \ (10^{-5} \text{ emu/g Oe}) \quad \text{(high field range)} \quad (4)$$
and
$$\chi_M = – 2.8 + 5\times10^{-7} \times (\mu_0 H)^{7.1} \ (10^{-5} \text{ emu/g Oe}) \quad \text{(spin-flipping field)} \quad (5).$$

The low field $\chi_M$ values are positive, and $\chi_M$ vs. $T$ appears to have an almost linear relationship. The high field $\chi_M$ values are negative, and $\chi_M$ vs. $T$ shows a complex relationship ($\propto T^{4.5}$). These features are very similar to those that appear in the superconducting state, and we emphasize that this RVB state may be the superconducting state. From Eq. (4), the superconducting critical temperature can be obtained as: $\chi_M \to 0$, $T_C^{SC} \approx 3.8$ K, and similarly, it can be obtained that the critical field of the emerging superconducting state is $H_C^{SC} \approx 8.92$ T. We noted that a similar non-conventional superconducting state was discovered in CeCoIn$_5$ heavy-Fermion alloy [49, 50]. It is seems that pairs of two aligned individual spins form RVB singlets as antiparallel-spin Cooper pairs and that this unconventional superconducting state is magnetically mediated superconductivity [51-53]. Another possible explanation is that the create local antiferromagnetism due to spin flipping at low temperature and high field rang, research into these phenomena is ongoing.

Although the above static susceptibility and magnetization measurements made it clear that the Néel long range ordering was suppressed, due to a combination of weak spin lattice frustration and field enhanced thermal and local spin fluctuations, the results showed that the field and thermal effects tuned the spins, causing the emergence of various kinds of exotic liquid-like, aligned spin, and spin flip or resonating-valence-bond (RVB) behaviors. The field tuning may go continuously to zero. However, when a spin-singlet (or RVB) phase with or without a gap emerges in response to spin excitations, it is a real problem how to distinguish between a weakly first order and a genuinely second order transition, and calculate critical exponents in the case of the latter [54]. Moreover, the likelihood that a local spin liquid model is appropriate is strengthened by the prediction of the RVB model of large entropy at low temperatures and a possible temperature ($T$) linear term in the heat capacity due to the electron spin density of states [46, 47]. The heat capacity is considered a very sensitive low-energy spectroscopic method for investigating low-energy excitations from the ground state. We can reliably explore what kind of ground state is realized through heat capacity measurements. In this respect, thermodynamic studies at low temperature are necessary and required for demonstrating the classical and quantum phases, especially the spin-liquid character of this material.

The temperature dependence of the heat capacities of single crystal Gd(OH)$_3$ nanowires under fields from 0 to 13.5 T are shown in Figure S4(a). The heat capacities, $C_P$, are plotted on a logarithmic scale against temperature. The $C_P$ show a slight upturn when the temperature is lower than 4 K for $H = 0$ T, and this is only 1.5 K lower than the value reported in Ref. [29] for single crystal data, which indicated that the Néel ordering temperature should be lower than $T_N \leq 0.94$ K. The figure also shows that heat capacities have slightly upturned shapes for $H < 2$ T curves, but no sharp peak structure for the $H \geq 2$ T curves. These results again indicate absence of long-range ordering over the entire measured temperature range for $H \geq 2$ T. We note that the heat capacity for the $H = 2$ T curve remains constant when the temperature T < 5 K. This may due to the coexistence of both a weak antiferromagnetic (AFM) phase and a quantum phase at this energy scale, which falls into the classical-quantum crossover regime, as discussed for the above susceptibility measurements. The absence of thermal anomalies when the system is under field $H \geq 2$ T is consistent with the above magnetic susceptibility measurements. To elucidate the detailed characteristics of low-energy excitations from the ground state, data in the low-temperature region are plotted as $C_P / T$ vs. $T^2$ in Figure 3(a). Fitting the 0T data between 6 and 10 K, using the formula

$$C_P / T = \gamma + \beta T^2 \qquad (5)$$

yields coefficients $\gamma = 51.52$ mJ K$^{-2}$ mol$^{-1}$ and $\beta = 1.51$ mJ K$^{-4}$ mol$^{-1}$. The finite electronic heat capacity coefficient $\gamma$ in this weakly frustrated quasi-1D $S = 7/2$ spin system suggests that gapless excitations occur from a local liquid-like ground state, similar to the case of κ-(BEDT-TTF)$_2$Cu$_2$(CN)$_3$ [55]. As Figure 3(b) shows, $\gamma$ is not seriously affected by magnetic fields within the finite temperature range where $C_P$ is linear with temperature up to 13.5 T. This excludes the possibility that paramagnetic impurity spins are the origin of the $\gamma$ term. The measurement was also performed on over 10 mg of single crystal nanowires, and thus, the large heat capacity at low temperatures is attributable to the intrinsic properties of this material.

As weak spin frustration combined with strong fluctuation spin systems may give rise to unexpected phase changes in the low energy region due to various kinds of many-body effects, such as in the hyperkagome system Gd$_3$Ga$_5$O$_{12}$, where the system appeared to have a field-induced transition from a spin-liquid phase to an antiferromagnetic phase [56]. For our system, it is obvious that the AFM phase is suppressed by field due to the increase in the frustration effects and enhancement of the fluctuation. However, the spin and the threshold field in that case are similar to those in the system studied here, suggesting that the physics here has a similar character. Moreover, Figure S4(a), Fig. 3(a), and Figure S4(b) shows a linear extrapolation of the data in the finite temperature range and then gives a vanishing $\gamma$ term in the heat capacity. Firstly, we analyse the linear extrapolation temperature range, which demonstrates that the $T$-linear contribution is observed only in the spin-liquid state. In heavy-electron systems produced by strong correlations of $f$-electrons in Ce and U compounds, anomalous divergence of $C_P/T$, expressed by the $-Log\ T$ term, is known as non-Fermi liquid behaviour at the quantum critical point [57], where the AFM phase is suppressed to $T_N = 0$ and an anomalous Fermi liquid phase appears. A similar kind of criticality might be expected in the critical region of a spin-liquid phase. In the case of this Gd(OH)$_3$ system, the low-temperature spin liquid state is realized below a crossover temperature of 20 K and H > 2 T, at which the heat capacity appears as a broad hump structure. As shown in Fig. 3(a) and Figure S4(b), a broad hump structure is observed in the $C_P/T^2$ vs. $T^2$ curve. Deviations of the experimental data of $C_P$ from $\beta T^3$ are determined using the formula:

$$\Delta C_P = C_P - \beta T^3, \qquad (6)$$

and $\Delta C_P/T$ values of these compounds are plotted as a function of temperature in Figure S4(b). The data clearly reveal the broad hump structure at peak temperature $T_{peak} = 26$ K. The magnitude of the $\Delta C_P/T$ peak is about 262.4 mJ /K /mol. As the lattice heat capacities are expressed by the simple $bT^3$ term around the hump temperature, the $\Delta C_P/T$ peak reflects the magnetic heat capacity in this compound. Similar methods have been used in single crystal heat capacity analysis [29] and have shown that this method more accurately estimates the lattice heat capacity in this compound. The hump temperature is consistent with the above susceptibility measurements, as it corresponds to the temperature of the spin fluctuation. Fig. 3(b) also reveals that the hump structure may be enhanced with magnetic field as it increases to 13.5 T.

The gapless or continuum gapped excitations observed in the heat capacity measurements can be discussed qualitatively by assuming finite density of states in spin excitations. The temperature dependence of the magnetic susceptibility has a broad peak around 26 K, which corresponds to strong fluctuation and a weakly frustrated spin chain lattice. Wilson's ratio has proved to be useful in characterising strongly correlated Fermi liquids. In a Fermi liquid, the low-temperature electronic specific heat $C_P$ is linear with temperature, with a slope $\gamma$. The magnetic susceptibility $\chi(T)$, is independent of temperature for low temperatures. Wilson's ratio (also known as the Sommerfeld ratio) is defined as the dimensionless quantity:

$$R_W \equiv 4\pi^2 k_B^2 \chi(0) / 3(g\mu_B)^2 \gamma \qquad (7)$$

where $g$ is the gyromagnetic ratio in the absence of interactions and $\mu_B$ is the Bohr magneton. For a non-interacting Fermi gas, $R_W = 1$. In terms of Landau's Fermi liquid parameters, $R_W = 1/(1+F_0^a)$. Wilson showed that for the Kondo model, the impurity contributions to $\chi(0)$ and $\gamma$ give a universal value of $R_W = 2$, independent of the strength of the interactions [58, 59]. From the analysis of inverse magnetic susceptibility, extrapolating the temperature dependence of the spin susceptibility down to $T = 0$ gives a residual magnetic susceptibility of $\chi(0)$ in the $H \geq 2$ T quantum regime or within the present spin-liquid temperature range. The $\chi(0)$ values are calculated from the above data on $1/\chi(0)$. Using the $\gamma$ value determined from the heat capacity in the linear – T range under different fields ($H \geq 2$T) in Fig. 3(a), the Wilson ratio, $R_W$, is evaluated to be 1.12, 1.05, 1.5, 2.1, and 2.16 for field of 2, 5, 7, 9, and 13.5 T, respectively, as shown in Fig. 3(b). $R_W$ has very similar values (~1) under fields from 2 T to 5 T, but increases to 1.5 under a 7 T field, and then shows similar values (~2) under fields from 9 to 13.5 T. These values of $R_W$ (1–2) imply a proper scaling of $\chi(0)$ and that $\gamma$ exists in the spin-liquid state. A

Wilson ratio on the order of unity is typical for a Fermi liquid with nearly free electrons, however, it shows that the magnetic field hardens electron spin, and reductions in temperature have the same effects as well. With decreasing temperature and increasing field, the increasing spin hardness or stiffness (for aligning spin) may enhance the spin interaction (dipolar interaction) from which spin flipping subsequently emerges, or spin hardening may increase the spin frequency, giving rise to the resonating-valence-bond state. These discussions are consistent with the susceptibility the inset of Fig. 1(a) and Fig. S3(d)), thermal magnetization (Fig. 2(b)) measurements, and moreover here, the heat capacity. As shown in Fig 3(a) and (b), there are inflexion points in the $Cp/T - T^2$ curve (Fig. S4(b)) or more clearly in the $Cp/T^2 - T$ curve (Fig. 3(b)), and we have denoted them as $T_{SF}^{HC}$. They are the points of emerging spin flipping and/or resonating-valence-bond behavior. The inflexion points $T_{SF}^{HC}$ are consistent with the above $T_C^*$ and $T_{SF}$ determined from susceptibility and magnetisation measurements, which will be discussed later.

All the phenomena discussed above may have benefits in terms of the magnetocaloric effect because they makes the spins in the system easier to align under external magnetic field, and especially in the spin liquid regime, the low lying energy and even gapless spectrum creates a high entropy ground state, while at the same time, the system can be easily put into a lowered entropy or lowered energy state and then may gain large entropy changes under external field, and this, plus the large spin potential (S = 7/2) of the $Gd^{3+}$ ions, may achieve very high MCE. Consequently we conducted a series of magnetic measurements and analysis the magnetocaloric effect properties in the $Gd(OH)_3$ nanowire system. Based on the local magnetic moment model, the greatest possible magnetic entropy change for the material would be (from the classical model):

$$\Delta S_M = R\, ln(2S+1) \approx 34.5 \text{ J / mol K}, \tag{8}$$

or in the quantum approach,

$$\Delta S_M = N\, k_B\, ln(2J+1) \approx 166 \text{ J / kg K}, \tag{9}$$

where R is the gas constant, S is the total spin angular momentum, N is the number of spins, J the quantum number of the spin, and $k_B$ the Boltzmann constant, which indicates that the system may achieve a large MCE in the low temperature range.

The magnetic entropy change $\Delta S_M$ was evaluated from the calculated isothermal magnetic entropy change $\Delta S_M(T, H)$ from the isothermal magnetization curves in the measured temperature range, according to Equation (10) below.

$$\Delta S_M(T,P,H)_{\Delta H, p} = \int_{H_1}^{H_2} \left(\frac{\partial M(T,P,H)}{\partial T}\right)_{H,p} dH \tag{10}$$

We now re-examine the data from the above isothermal magnetization curves of the $Gd(OH)_3$ nanowires, as shown in Fig. 3(a), which are plotted as a function of the applied magnetic field (ranging from $\mu_0 H = 0$ to 11 T) in the temperature range between 2.2 and 30 K. Under high magnetic field values (up to 11 T), the variation of the isothermal magnetisation (over the whole temperature range) becomes gradual, with a saturation being observed. We note that the peak magnetic moment reaches as high as 7.12 $\mu_B$ / $Gd^{3+}$ emu/g at 2.1 K and under field of 8.42 T, almost achieving the theoretical saturation moment of the $Gd^{3+}$ ion (~ 7.5 $\mu_B$).

According to Equation (10), the magnetic entropy changes ($-\Delta S_M$) can be calculated as a function of temperature for magnetic field changes from 0 – 1 to 0 – 11 T, and the $-\Delta S_M$ results are displayed in Fig. 4(a). The curves present a characteristic shape with no peak over the entire temperature range, indicating that there is no transition or no long range ordering of the Gd moments either, and the $-\Delta S_M$ appears to have no maximum and simply shows a linear increase with decreasing temperature. The maximum magnitude of $-\Delta S_M$ over the studied temperature range increases with increasing magnetic field change up to 27.8 J /kg K, 66 J / kg K, and 88 J / kg K at 2.55 K for field changes of 2 T, 5T, and 11 T, respectively. The magnetisation hysteresis loop at 2.1 K (not shown here) indicates that there is no hysteresis loss and also indicates fully reversible behaviour.

Employing the above heat capacity data (Fig. S4(a)) under external field from 0 to 13.5 T and in the temperature range from 2 to 15 K, we can calculate the adiabatic temperature change $\Delta T_{ad}$ (T, H) under external magnetic field change $\Delta H$ and at the temperature $T_0$ according to Equation (11) below:

$$\Delta T_{ad}(T_0, P, H)_{\Delta H, p} \cong T \frac{\Delta S_M(T_0, P, H)_{\Delta H, p}}{C_P(T_0, P, H)} \quad (11).$$

Since the heat capacity strongly depends on the external magnetic field, especially in the low temperature range [60], in order to relatively precisely calculate $\Delta T_{ad}(T, H)$, we re-calculated the $\Delta S_M(T, \Delta H)$ ($\Delta H$ = 0 to 1, 1 to 2, ···, 10 to 11 T) for every one Tesla magnetic field change, and the results are shown in Fig. S5(a). We then calculated the $\Delta T_{ad}(T, \Delta H)$ according to Eq. (11) using the measured $C_P$ data under relatively different external fields (as shown in Fig. S4(a)), and the results for $\Delta T_{ad}(T, \Delta H)$ are shown in Fig. S5(b). Finally, we can obtain the $\Delta T_{ad}(T, H)$ data under external field change of 0 to H, as shown in Fig. 4(b). The maximum adiabatic temperature change ($\Delta T_{ad}$) = 6.7, 17.6, and 36.4 K at 2.55 K for field changes of 2, 5, and 11 T, respectively. It is also interesting to note that the spin-flipping will generate a very large additional magnetocaloric effect in the system, as shown in Fig. 4(b). Moreover, the Gd(OH)$_3$ nanowires show extremely high $\Delta T_{ad}$, which satisfies one of the important criteria for selecting magnetic refrigerants, a large adiabatic temperature change.

The values of $-\Delta S_M$ obtained in the Gd(OH)$_3$ system are extremely large, with no magnetic/heat hysteresis, and are fully reversible. $-\Delta S_M$ is higher than in any reported reversible MCE magnetocaloric material (excluding the colossal or giant MCE materials with large hysteresis, as these materials normally exhibit confirmed maximum values of < 30 J/ kg K under a field change of 0 – 5 T at the ordering temperature). In comparison with the paramagnetic salts or nanosized garnets [21, 22], and well-designed molecular nanomagnets [20, 21] for ultra-low temperature application, which exhibit second order or first order paramagnetic to ferromagnetic (or antiferromagnetic) phase transitions at a few degrees Kelvin, the $\Delta S_M$ of the present material is 3 times higher than for those materials over the whole temperature range (2 to 20 K). In addition, a recent cytotoxicity investigation [61] has indicated that there are no cytotoxic effects from gadolinium hydroxide nanorods. These properties make Gd(OH)$_3$ nanorods a promising candidate as a refrigerant at ultra-low temperature.

Re-analyzing the magnetic entropy change ($\Delta S_M$) data by differentiating $\Delta S_M$ with magnetic field, we obtained the $\delta(\Delta S_M)/\delta(\mu_o H)$ values as functions of magnetic field at different temperatures. We define the $\delta(\Delta S_M)/\delta(\mu_o H)$ value as "$d\forall$" – the system's ability to support change in the magnetic entropy under finite environmental conditions (field and temperature):

$$d\forall = \delta(\Delta S_M) / \delta(\mu_o H)$$

(12).

Magnetic entropy ($S_M$) represents the degree of spin order in the spin system, while the magnetic entropy change ($\Delta S_M$) represents the degree of change in spin order, so that $d\forall$ represents the change in the velocity and strength of spin order changes under finite magnetic field changes.

Fig. S6(a) shows the magnetic entropy changes $-\Delta S_M$ at fields from 0 T to 11 T as a function of field for temperatures from 2.55 to 27.5 K. Differentiating the curves in Fig. S6(a), we can obtain $d\forall$ values at definite temperatures. Fig. S6(b) shows the $d\forall$ values at fields from 0 T to 11 T as a function of field for temperatures from 2.55 to 27.5K. The maximum value of $d\forall$ at a definite temperature ($T_s^*$) and field ($H_C^*$) can be determined, as such values appear at critical points and under boundary conditions. In fact, maximum $d\forall$ appears at the saturation field [62] of aligned spins near the saturated moment and the crossover points of the spins from classical to quantum behavior [63]. Fig. 5(a) shows the points of maximum value of $d\forall$ as a function of temperature (red triangle symbol line) and the boundary line (purple star symbol line) of aligned spins near the saturated moment temperature ($T_s^*$) and critical field ($H_C^*$) where the maximum value of $d\forall$ emerges. By fitting the two lines, the result can be obtained:

$$d\forall = 47.2 \times T^{-0.8}$$

(13)

for the red triangle symbol line, and

$$H_C^* = 1.3603 - 0.31 \times T_s^{*0.87}$$

(14)

From Eqs. (13) and (14), it can be seen that the $d\forall \to \infty$ and $H_C^* \approx 1.36$ T when the temperature $T \to 0$ K. We thus obtain the quantum critical point $H_{QC} \approx 1.36$ T, which is very consistent with above determined 1.4 T. At the quantum critical point, $d\forall \to \infty$, and this means that the system will generate enormous entropy changes when there is a small variation in the external field, according to Eq. (12), e.g., $d\forall \propto V_{eff}$, $\delta(\Delta S_M) \propto M_{eff}$, and $\delta(\mu_o H) \propto E_{eff}$. $V_{eff}$, $M_{eff}$, and $E_{eff}$ are the electron (spin) system's effective velocity, effective mass, and effective energy, respectively. The system will generate enormous velocity and strength in its variation of change in the degree of spin order when there are small changes in energy. Following Einstein's formula, $E = M \times C^2$, here it may need modify that the $d\forall \propto V_{eff}^2$, when the system effective mass $M_{eff} \to \infty$, we seem to be observing "black hole matter". In order to more clearly illustrate the image of the "black hole", we plot the 3-dimensional (3D) image of $d\forall$. Fig. 5(b) shows the 3D image of $d\forall$ as a function of field (from 0

T to 11 T) and temperature T (from 2.55 to 27.5 K). The shape of the 3D $d\forall$ plot in Fig. 5(b) is the cross-section at 2.55 K of the "black-hole". It will be closed at $T = 0$ K and it is a line when $d\forall \to \infty$ at the QCP. We note that some experiments have observed "zero point" entropy [64-66], as they show a macroscopic entropy at zero temperature. It has been predicted by anti de Sitter conformal field theory (*AdS-CFT*) that the black hole would be unstable and would eventually suck up "stuff" from its surroundings, covering its horizon with "hair" [5, 67]. In the electron spin system, out of the spin liquid state and at a quite low temperature, some unexpected order will set in that removes the ground-state entropy, giving it a unique ground state. Our experimental observations indicated that the spin-flip / fractional RVB / fractional superconducting-like state gives rise to a unique ground-state under ultra-cooled and high field conditions. Moreover, we may observe the above-mentioned "black hole" at the quantum critical point with $d\forall \to \infty$, for this understanding of the "black hole" is that when there is a small amount energy change (here, e.g., $\delta(\mu_oH)$ = small value), the system will generate enormous entropy change, and this will cause the system to suck up huge energy from its surroundings.

Following the above cross-checking of the magnetic measurements and analysis, we can plot a phase diagram, as shown in Fig. 6(a). The spin state boundary lines are determined from reference[29] and from the above thermodynamic susceptibility, heat capacity, magnetization, and magnetocaloric measurements and analysis. The AFM line only connects $T_N$ (the solid black square point) from Ref. [22] and the quantum critical point. The QCP was determined by analyzing susceptibility data $H_c^*$ in Fig. 1(a) and the magnetization data $H^*$ in Fig. S3(d). The red open circle line was determined by analyzing the susceptibility data $T_S$ in Fig. S2(e), and the purple solid star line was determined by analyzing the susceptibility data $T_S$ in Fig. S2(e). The solid blue square symbol line $T^*$ was determined from Fig. S2(b). The solid red circle line $T_{AS}$ and the open pink square line $T_{AS}^*$ were determined from Fig. S2(c) and S2(e). The solid green square line $T_C^*$ is from the inset of Fig. S2(b), the open blue diamond $T_{SF}$ line was determined from analyzing the magnetization curves of Fig. 2(a), and the open dark green star line is from Fig. 3(a). The above phase boundary lines were determined by different measurements, but they are very consistent with each other, which indicates that the methods we used confirm each other. Analyzing this phase diagram, we obtained the image of the electron spin interaction that is plotted in Fig. 6(b).

Fig. 6(b) presents a schematic diagram of the possible spin states and their relationships, quantum critical point, and quantum phase transitions. With decreasing temperature, the disordered spins (PM state) appear as a strong fluctuation (here, the peak temperature = 26 K) due to the quasi-1D spin chain structure, which is associated with strong thermal fluctuation. When the temperature is reduced to to the sub-Kelvin range (< 1 K), the weakly frustrated quasi-1D chains in the low field range spontaneously change to AFM ordering with the spin vector perpendicular to the chain. As the magnetic field (*H*) increases, *H* enhances the spin fluctuation and the thermal fluctuation, combined with weak frustration of the spin chain lattice structure, so that the AFM ordering is suppressed, and then there is emergence of the quantum critical point, exotic quantum phases, and quantum phase transitions due to strong correlation of the spins with increasing magnetic field. The exotic quantum phases that emerge include strong spin fluctuation, spin-liquid-like behavior, and an aligned spin state in the quantum critical regime (yellow (including light yellow) area in Fig. 6(b)) with enhancement of the strength of correlation of the spins. With $H$ increasing and $T$ cooling down, aligned spins in a high energy state are not stable, and "spin-flip" behavior emerges in the system, forming an RVB or even a superconducting-like state. We visualize the randomly ordered (or free direction) large spins (S = 7/2) with strong fluctuations as the "Spin Gas" region (light yellow), similar to the Fermi-gas regime. We visualize the state with short range ordered spins and aligned spins, but obviously not long-range ordered state, as the "Spin Liquid" region (yellow). We visualize the spontaneously ordered AFM state as the "Spin Solid I" (SS-I shown in figure) region (grey). We visualize the spin-flip or RVB ordered state as the "Spin Solid II" region (light grey). It is easy to see that the QCP separates the two solid spin states.

To summarize, high purity and highly uniform single crystal $Gd(OH)_3$ nanowires were synthesized by a facile hydrothermal method, and microstructural features were characterized by XRD, SEM, TEM, and HRTEM. Systematic magnetic susceptibility, magnetization, and heat capacity measurements, along with magnetocaloric analysis in magnetic fields up to 13.5 Tesla and temperatures down to 2 K indicated that, (1) the system presents strong fluctuation in its weak frustration and weakly antiferromagnetic nearest-neighbor quasi-1D chains; (2) the magnetic field enhances the thermal and local spin fluctuations, which suppress the long-range spin ordering (LRO), and LRO is not present within the measured temperature range in this system; (3) however, we observed possible field-induced, exotic spin-liquid-like, aligned-spin, and spin-flip states, exotic paramagnetic phases, and transitions in the low temperature and high field range, and determined the likely quantum critical point and quantum phase diagram; (4) we observed extremely large, fully reversible MCE (magnetic entropy change ($-\Delta S_M$) = 27.8, 66 and 88 J / kg K, adiabatic temperature change ($\Delta T_{ad}$) = 6.7, 17.6, and 36.4 K at 2.55 K for field changes of 2, 5, and 11 T, respectively, in this continuous range of quantum phase transitions in this system; (5) moreover, careful experiments and analysis may determine the quantum critical point (QCP) of this system; (6) analyzing the magnetocaloric effects close to the QCP, we get very large $d\forall = \delta(\Delta S_M) / \delta(\mu_o H)$ values, e.g. $d\forall \to \infty$ at the QCP, so it is possible to obtain large MCE and even exceed the theoretical value of MCE in the QCP system; (7) moreover, $d\forall \to \infty$ at the QCP indicates the existence of a "black hole"; (8) at high field, the spins show a spin-flipping phenomenon and may form resonating-valence-bond (RVB) singlets, and the magnetization and magnetocaloric effect measurements indicated that at this point, the emergence of RVB led to the observation of superconducting-like susceptibility behavior in this system.

The magnetic-entropy change monotonically increases with decreasing temperature, and this system exceeds the strength of the magnetocaloric effect (MCE) by 2 to 3 times in any other known low temperature reversible MCE material. The extremely large magnetic entropy change may be attributed to the large amount of weakly interacting spins that can be easily aligned in the low lying energy quantum critical regime and close to the QCP in our nanosized materials, since there is large MCE in the spin-liquid-like (low energy excitation and even gapless state) range. These indicate that the material is a promising MCE candidate for low temperature application, and possibly could make ultra-low temperatures easily achievable for most laboratories and for space application as well.

**Methods**

A single crystalline nanowire sample of Gd(OH)$_3$ was prepared by a facile hydrothermal method. The starting reagents, GdCl$_3$ (99.99%) and ammonia hydroxide (28-30%) solution, were purchased from Sigma-Aldrich. 0.6 g GdCl$_3$ was dissolved in 20 ml double-distilled water. Diluted NaOH solution was then added into the solution, and the pH was maintained at 9-10. The mixture was placed in a Teflon-lined autoclave and then heated up to 150 °C. This temperature was maintained for 24 h. The as-prepared sample was centrifuged and washed with distilled water 5 times, and then dried in vacuum oven at 50 °C. The x-ray diffraction (XRD) pattern was collected to confirm the crystallization on a GBC MMA diffractometer with Cu Kα radiation. The morphologies were characterized by using scanning electron microscopy (SEM, JEOL JSM-6460A) and transmission electron microscopy (TEM, JEOL 2010). The magnetization ($M$) and heat capacity ($C_P$) were measured under applied magnetic field of 0 – 13.5 Tesla (T), using the vibrating sample magnetometer (VSM) option of a Quantum Design 14 T Physical Property Measurement System (PPMS) in the temperature range of 2 – 300 K.

## Acknowledgements

The authors thank Dr. T. Silver for her help and useful discussions. This work is supported by the Australian Research Council.

# Figure Captions

**Figure 1 (a)** inverse susceptibility $(1/\chi) - T$ curves, which are upturned with increasing applied field; the inset is the plot of the $1/\chi - T$ curve with a 13.5 T field applied; **(b)** field dependence of inverse $T \rightarrow 0$ K susceptibility $(1/\chi_o)$, with the upper left inset showing $1/\chi - T$ curves for applied field from 200 Oe to 1 T and the lower inset showing $1/\chi - T$ curves for applied field from 2 T to 13.5 T.

**Figure 2 (a)** detailed magnetisation versus H curves for Gd(OH)$_3$ nanowires at temperatures from 2.1 K to 4.1 K; a magnified view of the low field region is presented in the left inset, and a magnified view of the high field region in the right inset; **(b)** low field and high field susceptibility determine the slope of the straight line parts of $M - H$ curves in the insets to Fig. 3(a), so the fitting lines and equations are presented for both the low and the high field ranges, respectively.

**Figure 3 (a)** $C_p / T$ vs. $T^2$ data in the low-temperature region; **(b)** $\chi(0) \approx \chi_0$ $(T \rightarrow 0$ K$)$, finite electronic heat capacity coefficient $\gamma$, and Wilson ratio $R_W$ determined by using low temperature data from 2.3 K to 15 K (see text) as a function of magnetic field ($H > 2$ T).

**Figure 4 (a)** Magnetic entropy change $-\Delta S_M$ (calculated from isothermal $M$-$H$ curves) of the Gd(OH)$_3$ single crystal nanowires at temperatures from 2.3 K to 30 K as a function of temperature for magnetic field changes from $0 - 1$ to $0 - 11$ T; **(b)** $\Delta T_{ad}$ at temperatures from 2.3 K to 30 K as a function of temperature for magnetic field changes from $0 - 1$ to $0 - 11$ T.

**Figure 5 (a)** points of maximum value of $\delta(\Delta S_M)/ \delta(\mu_o H)$ as a function of temperature (red triangle symbol line) and the boundary line (purple star symbol line) of aligned spin near the saturated moment temperature $(T_s^*)$ and the critical field $(H_C^*)$; at the maximum value of $\delta(\Delta S_M) / \delta(\mu_o H)$, $\delta(\Delta S_M) / \delta(\mu_o H) = 47.2 / T^{0.8}$ for the red triangle symbol line and $H_C^* = 1.3603 - 0.31$ T $\times T_s^*$ and by fitting the two lines, it can be seen that $\delta(\Delta S_M) / \delta(\mu_o H) \rightarrow \infty$ and $H_C^* \approx 1.36$ T when the temperature $T \rightarrow 0$ K; **(b)** 3-dimensional image of $\delta(\Delta S_M) / \delta(\mu_o H)$ as a function of field (from 0 T to 11 T) and temperature T (from 2.55 to 27.5 K).

**Figure 6** The spin state and quantum phase transitions diagram for the *Gd(OH)$_3$* nanowire system; **(b)** schematic diagram of possible spin states, their relationships, the quantum critical point, and the quantum phase transitions.

# Figures

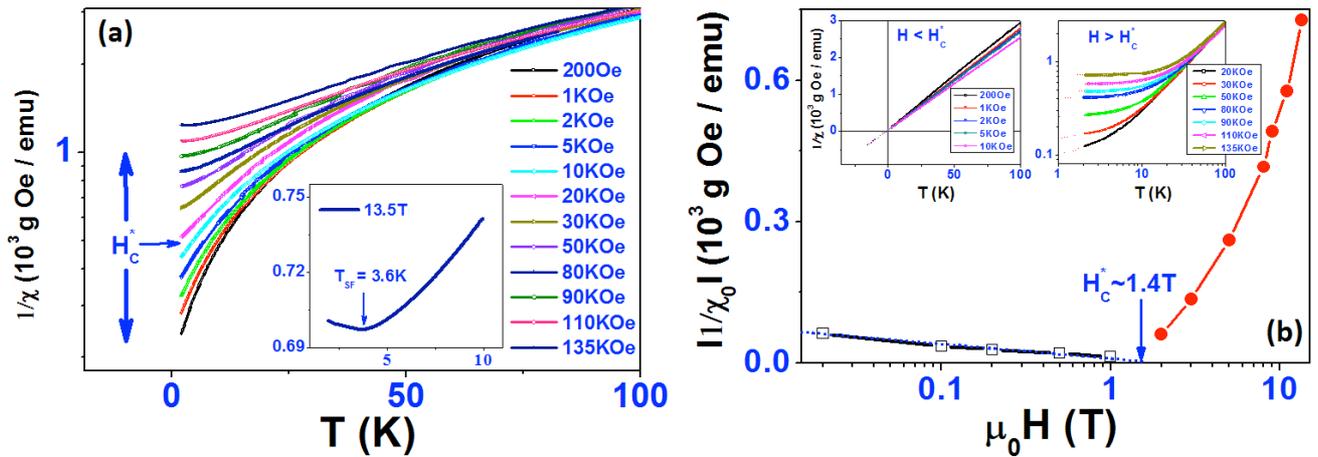

Fig. 1, R. Zeng et al.

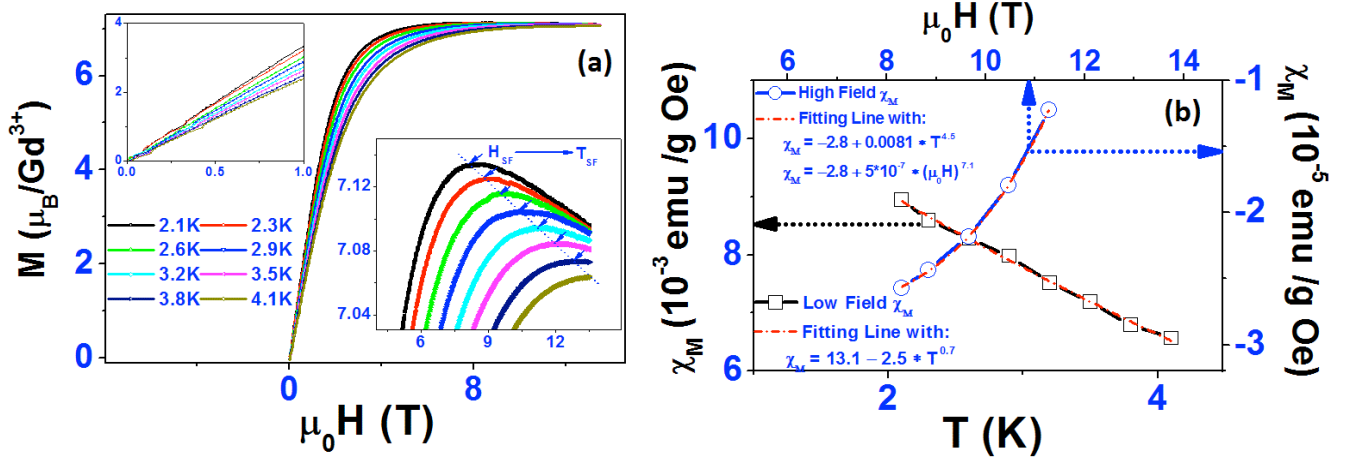

Fig. 2, R. Zeng et al.

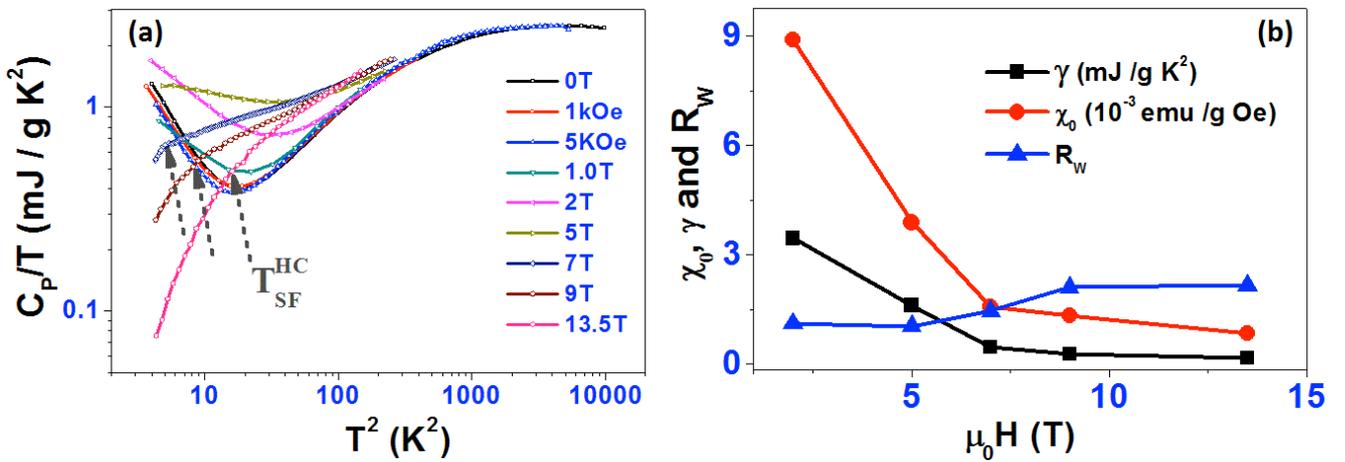

Fig. 3, R. Zeng et al.

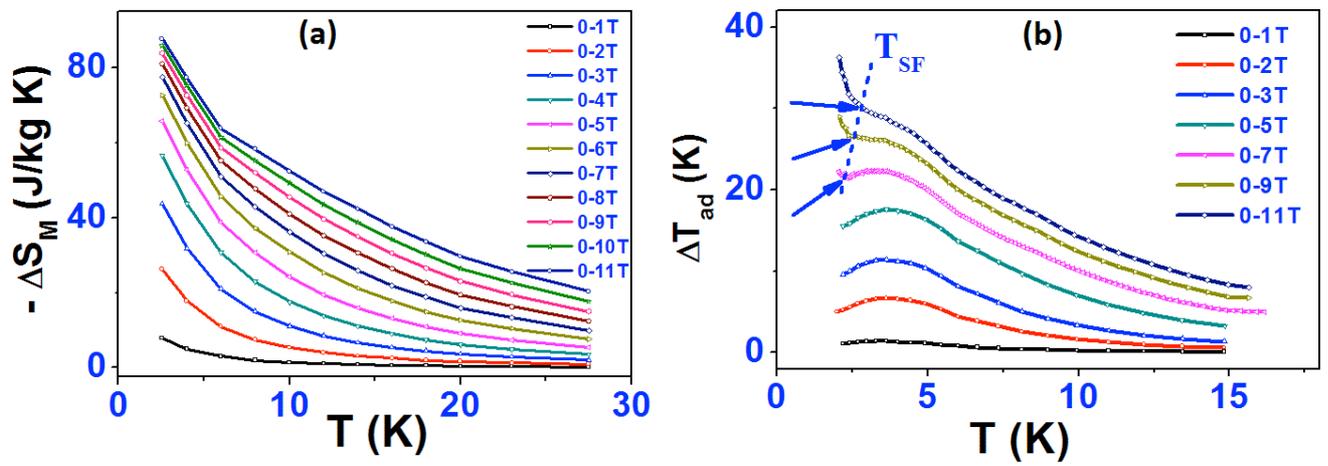

Fig. 4, R. Zeng et al.

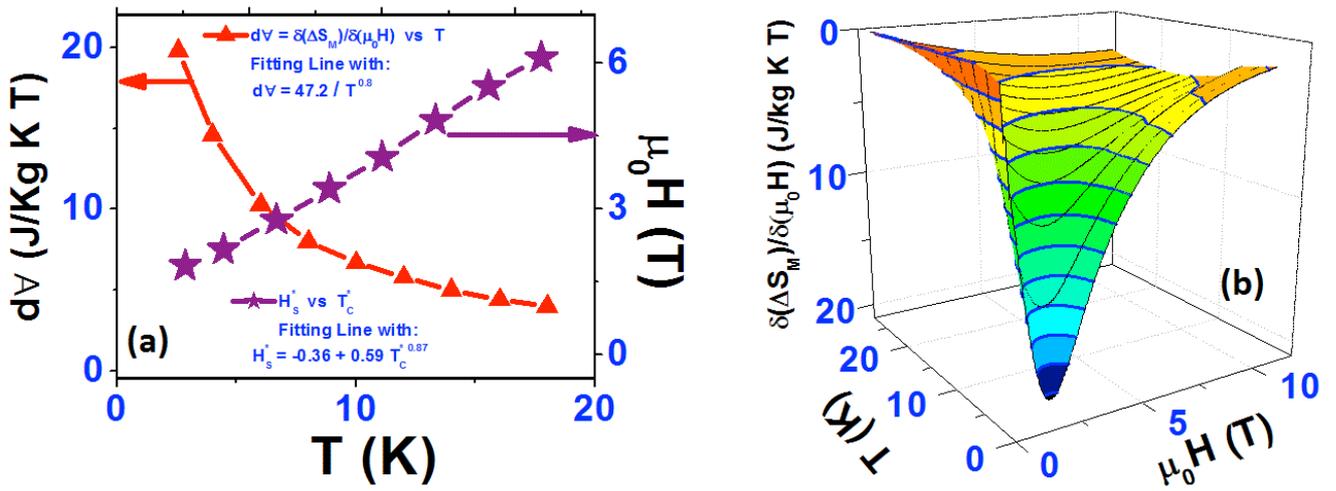

Fig. 5, R. Zeng et al.

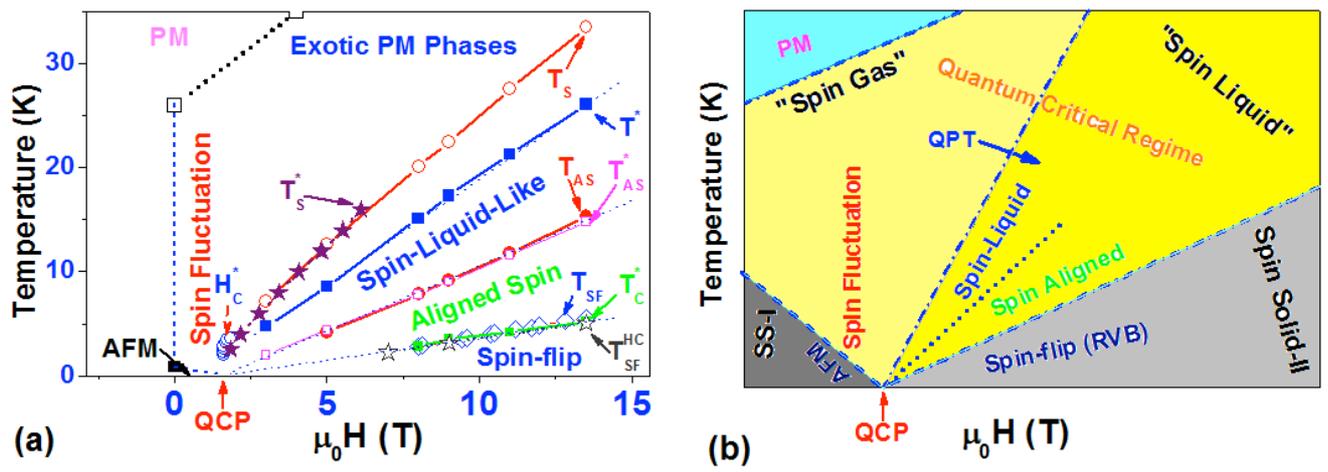

Fig. 6, R. Zeng et al.

# Supporting Information

# for

# Extremely Large Magnetic Entropy Changes, Quantum Phase Transitions and Diagram in Gd(OH)$_3$ Single Crystal Nanowires Quasi-One-Dimensional Large Spin (*S = 7/2*) Chain Antiferromagnet

R. Zeng [1,2], C.F. Zhang[1], J.C. Debnath[1], P. Shamba[1], J.L. Wang[1], Z.X. Chen[1], Z.P. Guo[1], S.X. Dou[1]

[1]*Institute for Superconducting and Electronic Materials, University of Wollongong, NSW 2522, Australia*

[2]*School of Computing, Engineering and Mathematics, University of Western Sydney, Campbelltown, NSW 2560, Australia*

Address for Correspondence:

Dr. R. Zeng

Solar Energy Technologies
School of Computing, Engineering and Mathematics
University of Western Sydney
Penrith Sout, Sydney, NSW 2751, Australia
Electronic mail: r.zeng@uws.edu.au

# I. Microstructure of highly uniformity Gd(OH)$_3$ nanowires:

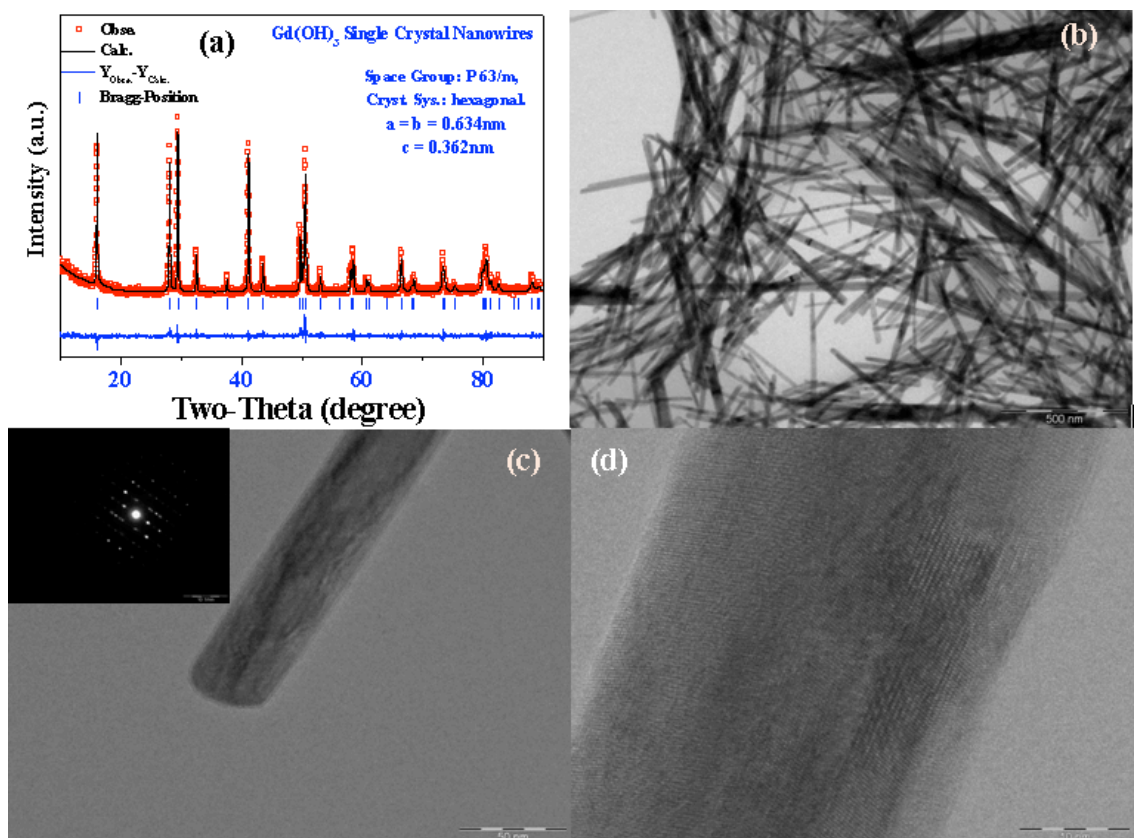

**Fig. S1 (a)** XRD pattern (symbols) and Rietveld refinement (solid line) of the Gd(OH)$_3$ nanowires; the short vertical lines mark the line positions in the standard, and the difference spectrum between experimental and calculated results is shown at the bottom; **(b)** FESEM image; **(c)** TEM image and SAED pattern (inset); and **(d)** HRTEM image of Gd(OH)$_3$ nanowires.

Fig. S1(a) displays the x-ray diffraction (XRD) and Rietveld refinement patterns of the Gd(OH)$_3$ nanowires. Fig. 1(b) is a field emission scanning electron microscope (FESEM) image of the Gd(OH)$_3$ nanowires, (c) contains a transmission electron microscope (TEM) image and the corresponding selected area electron diffraction pattern (SAED), and (d) shows a high resolution TEM (HRTEM) image of the Gd(OH)$_3$ nanowires. The above observations of the microstructure indicate that the nanomaterials show highly uniformity, high phase purity and are in single crystal nanowire form, with diameters of 25-35 nm and an average length of 800 nm. From the HRTEM image in Fig. 1(d), it appears that the nanowires are single crystals. The powder XRD refinement confirms the hexagonal crystal structure with space group *P 63/m* symmetry and lattice parameters *a* = *b* = 0.635 nm and *c* = 0.364 nm. In this structure, the Gd atoms form an array as infinite linear chains parallel to the *c*-axis. The distance between two neighboring Gd atoms is 0.364 nm, and each chain is surrounded by three other chains, so that they form hexagonal tunnels where the OH groups are inserted. The rings in the SAED pattern in the inset of Fig. 1(c) can be indexed as (100), (110), (200), (111), (201), and (002) reflections of hexagonal Gd(OH)$_3$; no extra reflections are observed, confirming that the phase is pure Gd(OH)$_3$.

## II. Susceptibility:

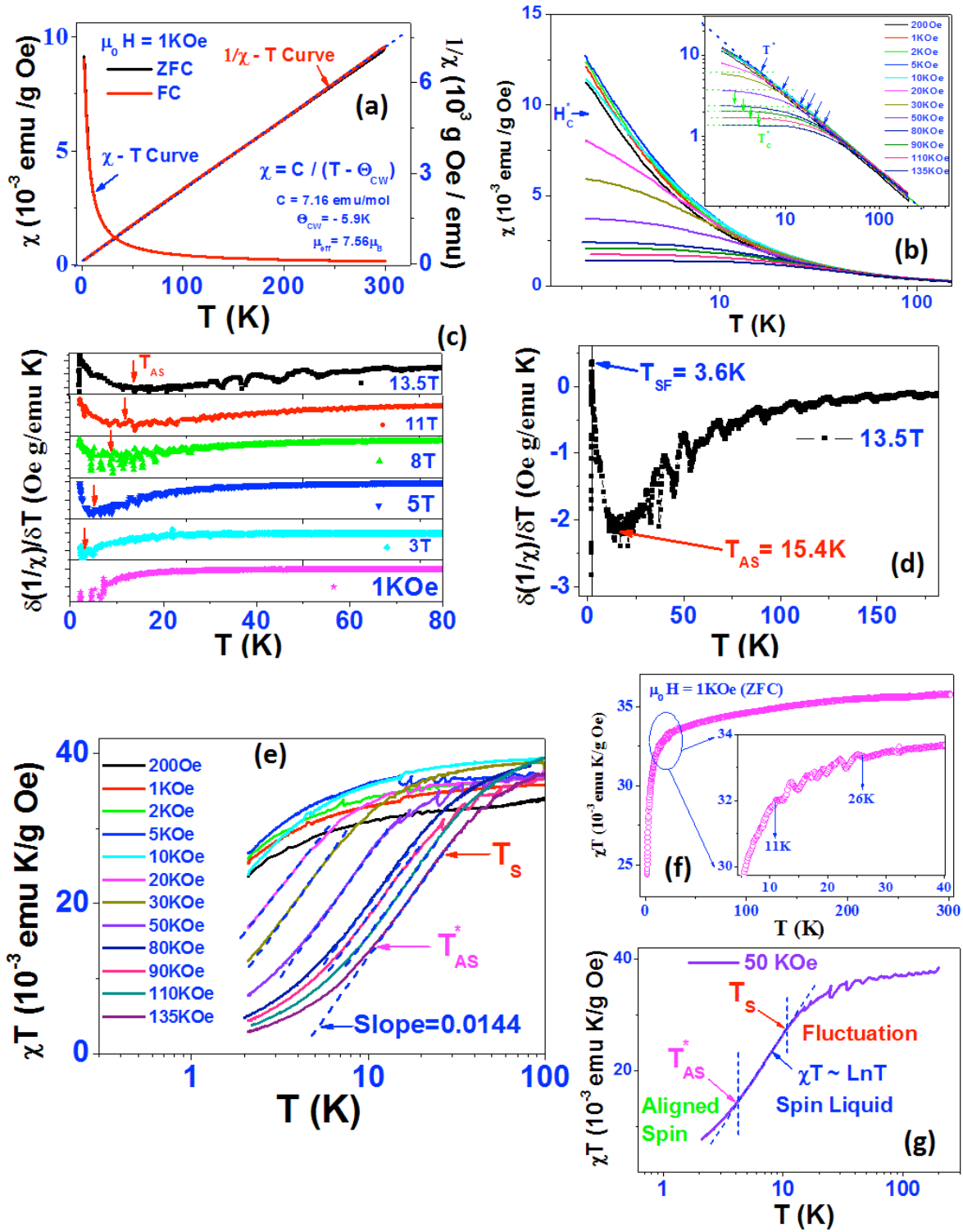

**Fig. S2 (a)** Temperature dependence of the magnetic susceptibility ($\chi$) and inverse susceptibility ($1/\chi$) under zero field cooling (ZFC: black symbol line) and field cooling (FC: red symbol line) in an external magnetic field $H = 1$ kOe ($\chi T - T$ and inverse $1/\chi - T$ curve), with the blue dashed line representing the Curie-Weiss fitting line; **(b)** $\chi T - T$ curves with the applied fields indicated; the inset shows log-log plots of $\chi T - T$ curves and the transition $T^*$ points are the crossover points of two extended lines (the blue dashed lines, the $\chi T - T$ curves of the paramagnetic state, and the green dashed lines, the $\chi T - T$ curves of another magnetic spin state), while $T^*_C$ indicates the points where the transition is fully completed; **(c)** $\delta(1/\chi)/\delta T - T$ curves for the indicated applied fields, allowing the transition points of the aligned spin state

($T_{AS}$) and the spin flip state ($T_{SF}$) to be determined; **(d)** $\delta(1/\chi)/\delta T - T$ curve for applied 13.5 T field, with the transition points at this field $T_{AS}$ = 15.4 K and $T_{SF}$ = 3.6 K; **(e)** temperature dependence of $\chi T$ under ZFC in an external magnetic field $H$ = 1000 Oe; the inset is an enlargement of the $\chi T - T$ curve at temperatures from 5 to 40 K; **(f)** temperature dependence of $\chi T$ under ZFC in an external magnetic field $H$ = 1000 Oe, with the inset showing an enlargement of the indicated region; **(g)** the temperature dependence of $\chi T$ under FC in an external magnetic field $H$ = 50000 Oe.

## III. Magnetization

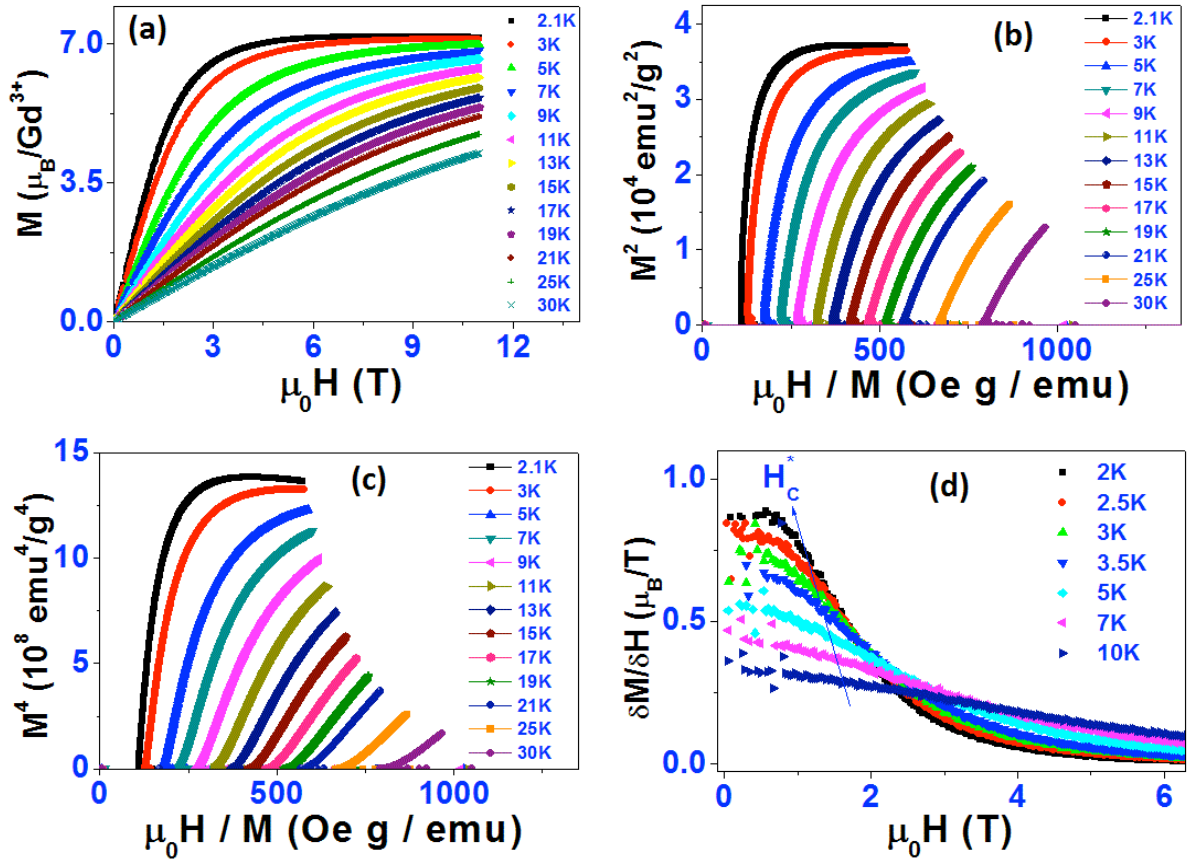

**Fig. S3 (a)** Field dependence of the magnetisation (M) measured at indicated temperatures from 2.1 K to 30 K; **(b)** Arrott-plot curves of $M^2$ versus $H/M$ for Gd(OH)$_3$ nanowires at temperatures from 2.1 K to 30 K; **(c)** Modified Arrott-plot curves of $M^4$ versus $H/M$ for Gd(OH)$_3$ nanowires at temperatures from 2.1 K to 30 K; **(d)** differentiated magnetisation ($\delta M/\delta H$) versus H curves for Gd(OH)$_3$ nanowires at temperatures from 2.1 K to 10 K.

## IV. Magneto-thermal properties:

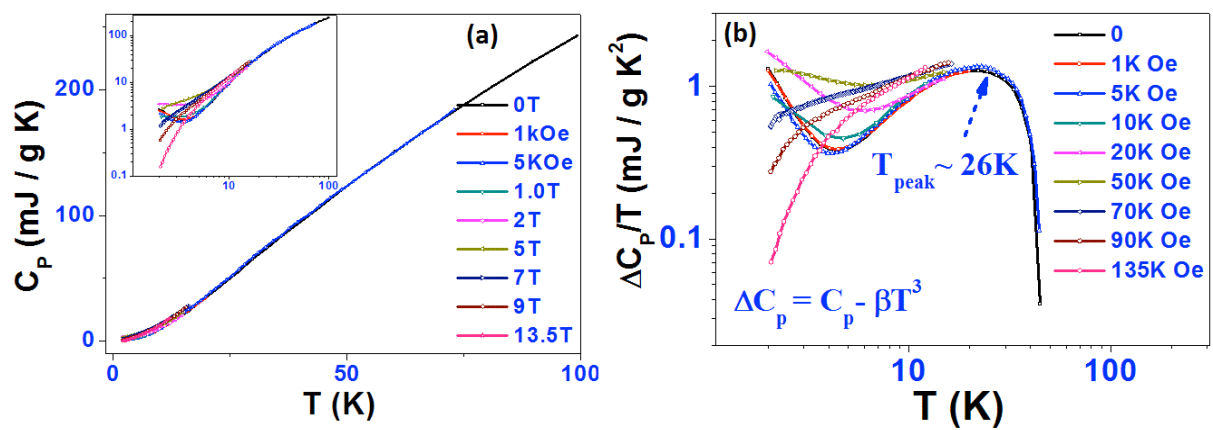

**Fig. S4 (a)** Temperature dependence of the heat capacities of single crystal Gd(OH)$_3$ nanowires under fields from 0 to 13.5 T, with the inset showing the log-log plots of heat capacities against temperature; **(b)** D$Cp/T$ values of these compounds as a function of temperature.

## V. Magnetocaloric effect (MCE):

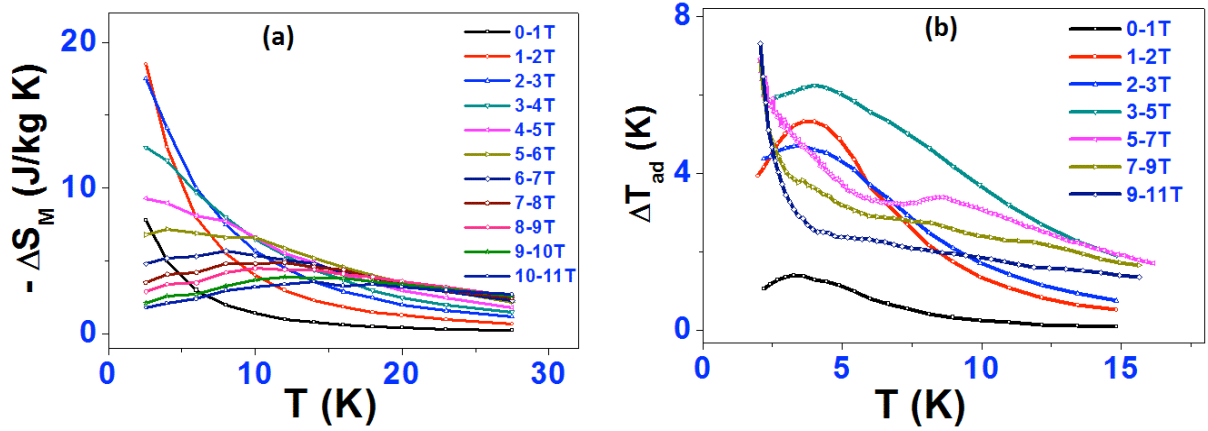

**Fig. S5 (a)** magnetic entropy changes $-\Delta S_M$ at temperatures from 2.3 K to 30 K as a function of temperature for magnetic field changes from 0 – 1 T, 1 – 2 T, up to 10 – 11 T, respectively; **(b)** $\Delta T_{ad}$ at temperatures from 2.3 K to 15 K as a function of temperature for magnetic field changes from 0 – 1, 1 – 2, 2 – 3, 3 – 5, 5 – 7, 7 – 9, and 9 – 11 T, respectively.

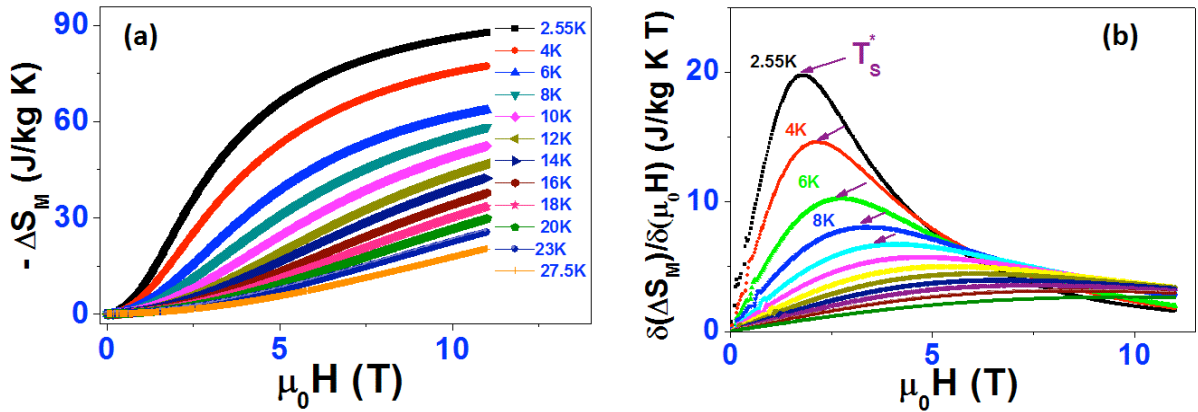

**Fig. S6 (a)** Magnetic entropy changes $-\Delta S_M$ at fields from 0 T to 11 T as a function of field for temperatures from 2.55 to 27.5 K; **(b)** $-\Delta S_M$ differentiated with magnetic field $[\delta(\Delta S_M)/\delta(\mu_0 H)]$ at fields from 0 T to 11 T as a function of field for temperature from 2.55 to 27.5 K.

## VI. Possible electron structure changes:

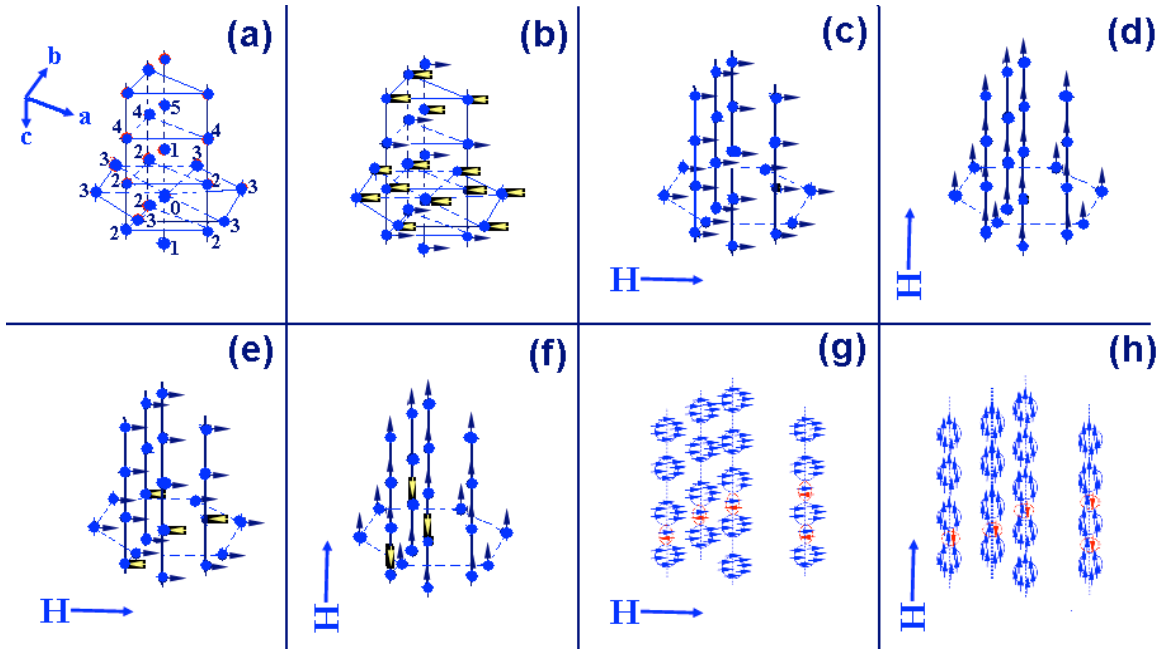

**Fig. S7 (a)** Arrangement of $Gd^{3+}$ ions in Gd(OH)$_3$: the reference ion is **0**, nearest neighbor ions **1n**, next nearest neighbor ions **2n**, and 3$^{rd}$, 4$^{th}$, 5$^{th}$ nearest neighbor ions are **3n**, **4n**, **5n**; the distances are 0.365 nm (= $c$), 0.408 nm, 0.634 nm, 0.655 nm, and 0.726 nm, respectively; **(b)** ideal arrangement of spins in the fully spontaneous magnetically ordered state; **(c)** ideal arrangement of aligned spins in the aligned spin state under high field perpendicular to the $c$-axis; **(d)** ideal arrangement of aligned spins in the aligned spin state under high field parallel to the $c$-axis; **(e)** ideal arrangement of aligned spins with a few flipped spins under high field parallel to the $c$-axis; **(f)** ideal arrangement of aligned spins with a few flipped spins under high field parallel to the $c$-axis; **(g)** ideal arrangement of aligned spins with a few flipped spins under high field perpendicular to the $c$-axis; **(h)** ideal arrangement of aligned spins with a few flipped spins under high field parallel to the $c$-axis.

We analyse the spin lattice microstructures and possible changes during transitions according to above experimental results. The arrangement of Gd$^{3+}$ ions in Gd(OH)$_3$ crystal is shown in Fig. S7(a), where the reference ion is **0**, the nearest neighbor ions **1n**, the next nearest neighbor ions **2n,** up to the 5$^{th}$ nearest neighbor ions, are located in the figure according to their distance. The ideal arrangement of spins occurs in the fully spontaneous AFM ordered state, and it and both spin aligned (parallel and perpendicular to the chain) states are shown in Fig. S7(b), (c), and (d), respectively. Under high field, aligned spins give rise to spin flipping or create a state with RVB pairs, and the states with the ideal arrangement of aligned spins and a few flipped spin under high field parallel and perpendicular to the spin chains are shown in Fig. S7 (e) and (f). Since $S = 7/2$, each arrow represents 7 single ½ spins. In the more detailed single spin arrangements of Fig. S7(g) and Fig. S7(h), in our experimental energy range (< 13.5 $T$), it may be that only two spins of each nearest neighbor ion are excited and appear to slip to create a RVB pair singlet. Our magnetization measurements may illustrate this point, since the moment value is slightly decreased with increasing field.